\documentclass[10pt,conference]{IEEEtran}

\usepackage{mathtools}
\usepackage{amsthm,amsmath,amssymb}
\usepackage{url}
\usepackage[ruled,vlined]{algorithm2e}
\usepackage{subfig,float}
\usepackage{tikz}
\usepackage{pgfplots}
\usepgfplotslibrary{patchplots}
\usepgfplotslibrary{groupplots}
\usepackage{tabularx}
\usepackage{booktabs}

\usepackage{pgf-umlsd}
\usepgflibrary{arrows}

\DeclareMathOperator*{\argmin}{arg\,min}

\begin{document}
\title{A Clustering-based Consistency Adaptation Strategy for Distributed SDN Controllers}
\author{
	\IEEEauthorblockN{Mohamed Aslan}
	\IEEEauthorblockA{Department of Systems and Computer Engineering\\ Carleton University, Ottawa, ON K1S 5B6, Canada.\\
		Email: maslan@sce.carleton.ca}
	\and
	\IEEEauthorblockN{Ashraf Matrawy}
	\IEEEauthorblockA{School of Information Technology\\ Carleton University, Ottawa, ON K1S 5B6, Canada.\\
		Email: ashraf.matrawy@carleton.ca}
}

\maketitle

\begin{abstract}
Distributed controllers are oftentimes used in large-scale SDN deployments where they run a myriad of network applications simultaneously.
Such applications could have different consistency and availability preferences. 
These controllers need to communicate via east/west interfaces in order to synchronize their state information.
The consistency and the availability of the distributed state information are governed by an underlying consistency model.
Earlier, we suggested \cite{aslan2016adaptive} the use of adaptively-consistent controllers that can autonomously tune their consistency parameters in order to meet the performance requirements of a certain application.
In this paper, we examine the feasibility of employing adaptive controllers that are built on-top of tunable consistency models similar to that of Apache Cassandra.
We present an adaptation strategy that uses clustering techniques (sequential k-means and incremental k-means) in order to map a given application performance indicator ($\chi$) into a feasible consistency level ($\Phi$) that can be used with the underlying tunable consistency model.
In the cases that we modeled and tested, our results show that in the case of sequential k-means, with a reasonable number of clusters ($\geq$ 50), a plausible mapping (low RMSE) could be estimated between the application performance indicators ($\chi$) and the consistency level indicator ($\Phi$). In the case of incremental k-means, the results also showed that a plausible mapping (low RMSE) could be estimated using a similar number of clusters ($\geq$ 50) by using a small threshold ($\simeq$ 0.01). 
\end{abstract}


\IEEEpeerreviewmaketitle

\section{Introduction}\label{sec:intro}
Recent research \cite{dixit2013towards, tootoonchian2010hyperflow, levin2012logically, koponen2010onix, berde2014onos} in Software-Defined Networking (SDN) employs multiple distributed controllers for scalability and reliability reasons.
Using distributed controllers allows the network to scale-out without introducing bottlenecks or single point of failure.
It also provides the network with redundancy and fault-tolerance.

Distributed SDN controllers need to communicate (via east/west interfaces) in order to synchronize their state information (we call this process \emph{controller state distribution}). Hence, they are subjected to issues similar to those affecting distributed datastores \cite{panda2013cap}.
A major issue is the trade-off between consistency and availability in the case of network partitioning, which was identified by Eric Brewer in the CAP (Consistency, Availability and Partitioning) conjecture \cite{brewer2000towards, brewer2012cap}.
The CAP conjecture states that in the case of network partitioning, a distributed system will have to choose between the consistency of the data or the availability of the system. Systems that prefer consistency over availability are labeled as \emph{strongly-consistent} systems. While, systems that have the ability to change their behavior (degree of consistency) are known as \emph{tunably-consistent} systems \cite{cassandraconfig, yu2000building}.

In SDN, the consistency level of state information exchanged among the distributed controllers can negatively affect the network application performance \cite{levin2012logically, Guo201495, aslan2016impact}, depending on the performance indicators being considered.

There exist a multitude of SDN applications, having different performance indicators. As such, some of these applications can tolerate the state information inconsistency for the sake of higher availability. Therefore, applications could be built on-top of tunably-consistent distributed controllers which could be tuned differently for each application. Earlier \cite{aslan2016adaptive}, we proposed the use of adaptive controllers \cite{aslan2016adaptive} running on-top of tunably-consistent controllers in order to autonomously handle setting the parameters of the tunably-consistent distributed controllers based on application-specific performance indicators.

In this paper, we investigate the feasibility of using adaptive controllers running on-top of tunable consistency models similar to that of Apache Cassandra \cite{lakshman2009cassandra, lakshman2010cassandra} or Amazon DB \cite{sivasubramanian2012amazon}.
We present a controller adaptation strategy that - given an application-specific indicator ($\chi$) - autonomously tunes the consistency level ($\Phi$) of the distributed controllers in order to maintain a certain value for such application-specific indicator. In presenting such strategy, we make the following contributions:
(1) we show how to quantize the level of consistency (subsection \ref{sec:proposed:tcm}) and how to use it in selecting appropriate values for the tunable consistency model parameters, and
(2) we show how online clustering techniques can be employed (subsection \ref{sec:proposed:am}) in order to map the application-specific performance indicators ($\chi$) into various consistency levels ($\Phi$).

The rest of this paper is organized as follows:
In \S\ref{sec:discussion}, we discuss the need for adaptive controllers in distributed SDN deployments.
We provide an overview on the topics of eventual and tunable consistency models in \S\ref{sec:background}.
\S\ref{sec:proposed} is the proposed realization of adaptive SDN controllers.
The evaluation is presented in \S\ref{sec:results}.
Finally \S\ref{sec:conclusion} will be our conclusion and an outline for possible foreseeable work.

\section{The Case for SDN Adaptive Controllers}\label{sec:discussion}
As aforementioned, the use of distributed controllers in large-scale SDN deployments is crucial. First, it can reduce the control delays as the control load is now handled by multiple controllers opposed to a single one. Second, it extricates the network from having a single point of failure embodied by the controller, hence increases the reliability and fault-tolerance of the network. Finally, employing distributed controllers allows the network to scale-out (horizontally) by adding more controllers. Dixit \emph{et al.} \cite{dixit2013towards}, suggested dynamically growing and shrinking the pool of controllers based on the traffic conditions, and to get rid of the controller/switch static mapping which can led to uneven distribution of the control load.

Managing distributed controllers in large-scale SDN environments can be a daunting task. First, those controllers are subjected to issues that affect distributed datastores \cite{panda2013cap} including the trade-off between consistency and availability of data during network partitioning. Next, there is a great number of SDN applications different in their requirements and employ different performance indicators. As such, some SDN applications may prefer different consistency and availability configurations \cite{koponen2010onix}. Finally, two or more applications with different requirements (could be conflicting) might be running on the controllers at the same time.

An example for an application that might prefer to lower its consistency for higher availability - as long as it is maintaining a certain level of performance - would be a load-balancer. The load-balancer would need to maintain information about the current load distribution in the network. However, as long as it is not creating routing loops (more in \cite{Guo201495}), it can tolerate some inconsistency in order to achieve a higher degree of availability. On the other hand, a firewall might represent an application that would not tolerate inconsistency and would prefer to be strongly consistent at the expense of being available.

We believe that distributed controllers that employ a tunable consistency model (similar to that of Apache Cassandra; see subsection \ref{sec:background:model}) are more suitable for large-scale SDN deployments that simultaneously run myriad of heterogeneous network applications.
Onix \cite{koponen2010onix} lets the applications make their own trade-off between consistency and availability by providing them with two data-stores: (1) a strongly consistent transactional data-store, and (2) an eventually consistent (more in the next section) in-memory distribute hash table (DHT).

Furthermore, we believe that an effective strategy to handle the case of heterogeneous applications, is by extending tunable consistency with an adaptive mode.
In such mode, the controllers given a per-application performance indicator will monitor the network behavior and adapt to the current conditions by autonomously tuning their consistency levels \cite{aslan2016adaptive}.
Adaptive distributed controllers can reduce the SDN applications development and maintenance cost by shifting the complexity of handling distribution issues out of the applications, reducing the application complexity. In addition, they can reduce the overhead of state distribution among the controllers.

\section{Background on Consistency}\label{sec:background}
In this section, we explain the consistency model used in a number of modern data-stores such as Apache Cassandra \cite{lakshman2009cassandra, lakshman2010cassandra} and Amazon DynamoDB \cite{sivasubramanian2012amazon}.

\subsection{Notations}
In distributed controllers, data are copied and stored at different controllers, such copies are known as \emph{replicas}. In this paper, we assume that no more than a copy of a certain data item will be stored at the same controller. We also use the term \emph{replicas} when referring to the machines storing the data copies.
Table \ref{tab:notation} shows the notations used throughout this paper.

\begin{table}[h!]
\caption{Notations used in this paper.}
\begin{tabularx}{.5\textwidth}{c | X}
	\toprule
	Symbol & Definition\\
	\midrule
	$M$ & the total number of nodes in a controllers cluster.\\
	$N$ & the number of replicas ($N \leq M$), assumed to be set based on network policy and hence constant.\\
	$R$ & the number of replicas that must confirm the read operation in order to be successful ($1 \leq R \leq M$).\\
	$W$ & the number of replicas that must confirm the write operation in order to be successful ($1 \leq W \leq M$).\\
	$\Phi$ & the consistency level indicator at the controller.\\
	$\chi$ & the application-specific performance indicator.\\
	\bottomrule
\end{tabularx}
\label{tab:notation}
\end{table}

\subsection{The Tunable Consistency Model}\label{sec:background:model}
The consistency model employed by Apache Cassandra is both an eventual and a tunable consistency model. Eventual consistency \cite{bailis2012probabilistically} is a consistency model where all replicas eventually receive the most up-to-date values after sometime if no further updated occurred. With tunable consistency, we refer to a property of a consistency model where the level of consistency can manually be tuned.
Cassandra allows the application to select between a number of predefined consistency levels, the most releavant ones are: (1) ONE, (2) QUORUM, and (3) ALL \cite{cassandraconfig}.
The first level `ONE' indicates that an operation is considered successful if one replica ($R = 1$) returned the most recent version in case of a read operation, or a confirmation is received from one replica ($W = 1$) in case of a write operation. This level provides a low latency and a high availability.
The second level `QUORUM' ($R + W > N$) indicates that an operation is considered a success if a quorum of replicas returned the most up-to-date version in case of a read operation, or a confirmation is received from quorum of replicas in case of a write operation. This level ensures strong consistency.
Finally, the `ALL' level indicates that an operation is considered a success if all of the replicas ($R = N$) responded and the most up-to-date version is calculated in case of a read operation, or a confirmation is received from all of the replicas ($W = N$) in case of a write operation. This level provides highest possible consistency level but the lowest availability.

\begin{figure}
\centering
\subfloat[Case I: $W=3$, $R=3$ (strong consistency)]{
	\label{tunable:a}
	\resizebox{0.5\textwidth}{!}{
	\begin{tikzpicture}
\tikzstyle{vertex}=[circle,fill=black!25,minimum size=18pt,inner sep=0pt]
\tikzstyle{selected}=[vertex,draw=black,dashed,thick]
\tikzstyle{doubleseleted}=[vertex,fill=black,text=white]

% initially
\begin{scope}
	% none
	\foreach \name/\angle/\text in {V-1/234/c_{1}, V-2/162/c_{2}, V-3/90/c_{3}, V-4/18/c_{4}, V-5/-54/c_{5}}
		\node[vertex,xshift=6cm,yshift=.5cm] (\name) at (\angle:1cm) {$\text$};
	% connections
	\foreach \from/\to in {1/2,2/3,3/4,4/5,5/1,1/3,2/4,3/5,4/1,5/2}
    	\draw (V-\from) -- (V-\to);
    % label
	\node[inner sep=0pt,label={60:$t_{0}$}] at (5.5,-1.5) {};
\end{scope}

% write
\begin{scope}[shift={(3,0)}]
	% selected
	\foreach \name/\angle/\text in {W-1/234/c_{1}, W-2/162/c_{2}, W-3/90/c_{3}}
		\node[selected,xshift=6cm,yshift=.5cm] (\name) at (\angle:1cm) {$\text$};
	% none
	\foreach \name/\angle/\text in {W-4/18/c_{4}, W-5/-54/c_{5}}
		\node[vertex,xshift=6cm,yshift=.5cm] (\name) at (\angle:1cm) {$\text$};
	% connections
	\foreach \from/\to in {1/2,2/3,3/4,4/5,5/1,1/3,2/4,3/5,4/1,5/2}
    	\draw (W-\from) -- (W-\to);
    % label
	\node[inner sep=0pt,label={60:$t_{1}$: write}] at (5,-1.5) {};
\end{scope}

% read
\begin{scope}[shift={(6,0)}]
	% none
	\foreach \name/\angle/\text in {RW-1/234/c_{1}, RW-2/162/c_{2}}
		\node[vertex,xshift=6cm,yshift=.5cm] (\name) at (\angle:1cm) {$\text$};
	% selected
	\foreach \name/\angle/\text in {RW-4/18/c_{4},RW-5/-54/c_{5}}
		\node[selected,xshift=6cm,yshift=.5cm] (\name) at (\angle:1cm) {$\text$};
	% double
	\foreach \name/\angle/\text in {RW-3/90/c_{3}}
		\node[doubleseleted,xshift=6cm,yshift=.5cm] (\name) at (\angle:1cm) {$\text$};
	% connections
	\foreach \from/\to in {1/2,2/3,3/4,4/5,5/1,1/3,2/4,3/5,4/1,5/2}
    	\draw (RW-\from) -- (RW-\to);
    % label
	\node[inner sep=0pt,label={60:$t_{2}$: read}] at (5,-1.5) {};
\end{scope}

\end{tikzpicture}
	}
}\\[5pt]
\subfloat[Case II: $W=3$, $R=2$ (eventual consistency)]{
	\label{tunable:b}
	\resizebox{0.5\textwidth}{!}{
	\begin{tikzpicture}
\tikzstyle{vertex}=[circle,fill=black!25,minimum size=18pt,inner sep=0pt]
\tikzstyle{selected}=[vertex,draw=black,dashed,thick]
\tikzstyle{doubleseleted}=[vertex,fill=black,text=white]

% initially
\begin{scope}
	% none
	\foreach \name/\angle/\text in {V-1/234/c_{1}, V-2/162/c_{2}, V-3/90/c_{3}, V-4/18/c_{4}, V-5/-54/c_{5}}
		\node[vertex,xshift=6cm,yshift=.5cm] (\name) at (\angle:1cm) {$\text$};
	% connections
	\foreach \from/\to in {1/2,2/3,3/4,4/5,5/1,1/3,2/4,3/5,4/1,5/2}
    	\draw (V-\from) -- (V-\to);
    % label
	\node[inner sep=0pt,label={60:$t_{0}$}] at (5.5,-1.5) {};
\end{scope}

% write
\begin{scope}[shift={(3,0)}]
	% selected
	\foreach \name/\angle/\text in {W-1/234/c_{1}, W-2/162/c_{2}, W-3/90/c_{3}}
		\node[selected,xshift=6cm,yshift=.5cm] (\name) at (\angle:1cm) {$\text$};
	% none
	\foreach \name/\angle/\text in {W-4/18/c_{4}, W-5/-54/c_{5}}
		\node[vertex,xshift=6cm,yshift=.5cm] (\name) at (\angle:1cm) {$\text$};
	% connections
	\foreach \from/\to in {1/2,2/3,3/4,4/5,5/1,1/3,2/4,3/5,4/1,5/2}
    	\draw (W-\from) -- (W-\to);
    % label
	\node[inner sep=0pt,label={60:$t_{1}$: write}] at (5,-1.5) {};
\end{scope}

% read
\begin{scope}[shift={(6,0)}]
	% none
	\foreach \name/\angle/\text in {RW-1/234/c_{1}, RW-2/162/c_{2}, RW-3/90/c_{3}}
		\node[vertex,xshift=6cm,yshift=.5cm] (\name) at (\angle:1cm) {$\text$};
	% selected
	\foreach \name/\angle/\text in {RW-4/18/c_{4},RW-5/-54/c_{5}}
		\node[selected,xshift=6cm,yshift=.5cm] (\name) at (\angle:1cm) {$\text$};
	% connections
	\foreach \from/\to in {1/2,2/3,3/4,4/5,5/1,1/3,2/4,3/5,4/1,5/2}
    	\draw (RW-\from) -- (RW-\to);
    % label
	\node[inner sep=0pt,label={60:$t_{2}$: read}] at (5,-1.5) {};
\end{scope}

\end{tikzpicture}
	}
}
\caption{Tunable Consistency Model in $\mathcal{O} (1)$ P2P Distributed Datastores}
\label{tunable}
\end{figure}
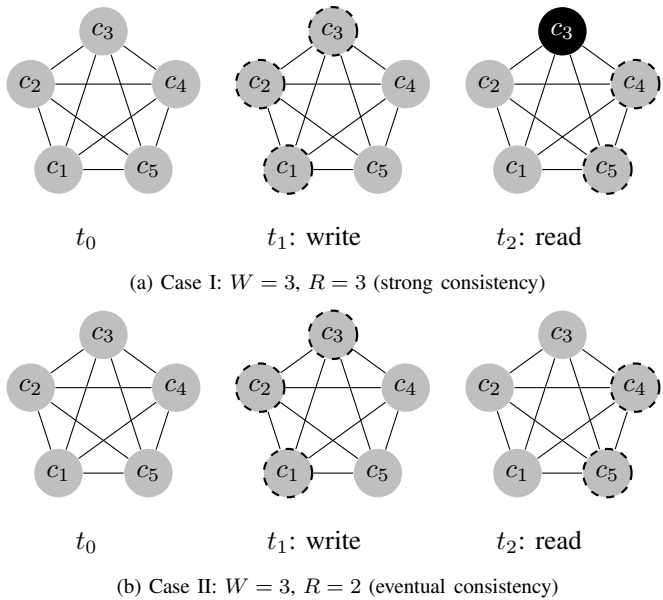

For example (Fig. \ref{tunable}), for a write (or update) operation to be succeed it must be written successfully on $W$ different nodes, and for a following read operation to succeed $R$ nodes must respond and return some value.
In the first case (Fig. \ref{tunable:a}), $N=5$, $W=3$, and $R=3$.
At $t_{1}$, a write operation was requested and confirmed by three nodes (at random): $c_{1}$, $c_{2}$, and $c_{3}$, while the operation might have failed at $c_{4}$ and $c_{5}$, the overall operation is marked a success (recall $W=3$).
At $t_{2}$, a read operation was initiated and only three nodes (at random): $c_{3}$, $c_{4}$, and $c_{5}$ returned an answer.
And since $R + W > N$ then for sure one node from those that answered the read operation will hold the most up-to-date value, in this example it is node $c_{3}$.
On the other hand, in the second case (Fig. \ref{tunable:b}), $N=5$, $W=3$, and $R=2$.
At $t_{2}$, only two nodes (at random): $c_{4}$, and $c_{5}$ returned an answer to the read operation, yet the overall read operation is marked a success (recall $R=2$). Those nodes may not have the most up-to-date value. Thus, there is no guarantee if $R + W \le N$ that a read operation will return the most up-to-date value. However, after sometime, if no further updates occurred, all nodes will \emph{eventually} receive the most up-to-date values.

As aforesaid, controllers might simultaneously be running multiple network applications, each having its own requirements.
The number of replicas ($N$) could also be application-specific, \emph{e.g.,} an application dealing with more important information would choose a higher number of replicas whereas an application dealing with less important information would choose a less value for $N$.
Even though the number of replicas ($N$) is application-specific, the nodes (controllers) themselves that are responsible for maintaining such replicas are decided by a consistent hashing function \cite{lakshman2010cassandra}.

\section{Proposed Adaptation Strategy}\label{sec:proposed}
The adaptation strategy requires the collaboration of different modules of an adaptive controller (proposed in \cite{aslan2016adaptive}).
In this section, we describe some of the modules needed for realizing the adaptive controllers architecture, namely: (1) the stored procedure compiler module, (2) the tunable consistency module, and (3) the adaptation module.

\subsection{Stored Procedure Compiler Module}
Applications often require different performance indicators ($\chi$). This module is needed to allow applications to instruct the controllers how to calculate their performance indicators (e.g., standard deviation between the loads in case of a load-balancing application). Moreover, in case that the applications are to run on physically separate machines from the controllers, the task of calculating the performance indicators is shifted to the controllers to reduce delays caused by the applications-controllers communication. An application installs a stored procedure similar to that used in Database systems \cite{iso:sql} at the controller which can be executed by the stored procedure compiler module in order to calculate the value of the application-specific performance indicator ($\chi$), when needed.
We assume that security measures are taken to prevent exploiting the use of the stored procedure compiler module and to ensure safe execution of the stored procedures at the controllers. The security aspects of the controllers are outside the scope of this paper.

\subsection{Tunable Consistency Module}\label{sec:proposed:tcm}
This module provides the adaptation module with a configurable consistency level parameter ($\Phi$) that can be tuned in order to change the level of consistency.

\noindent \textbf{Consistency Level Parameter.}
As aforesaid, the adaptation module requires a parameter that can be tuned in order to change the consistency level.
In the proposed strategy, we adopt the tunable consistency model discussed in section \ref{sec:background:model} as a base for our tunable consistency module.
Such model provides $R$, $W$, and $N$ as configurable parameters. However, mapping those parameters to a performance indicator ($\chi$) could be complex for the adaptation module.
$R$, $W$, and $N$ are specific parameters to this particular consistency model (Cassandra-like). Therefore, exposing $R$, $W$, and $N$ to the adaptation module would lower the modularity of the system \emph{i.e.,} it will be harder to replace the tunable consistency module with another one without having to modify the adaptation module. 
Hence, the tunable consistency module provides the adaptation module with a single tunable parameter ($\Phi$) that directly relates to the consistency level, and the tunable consistency module is responsible for mapping that parameter ($\Phi$) into its internal specific parameters (\emph{e.g.,} $R$, $W$, and $N$).

\begin{figure}[h]
\centering
\begin{tikzpicture}
	\begin{axis}[%
		%3d box,
		colorbar,
		colorbar style={
			width=0.15cm
		},
		%view={25}{25},
		width=4.5cm, height=4.5cm,
		scale only axis,
		xlabel={$R$}, ylabel={$W$}, zlabel={$\Phi$},
		xmin=1, xmax=20, xmajorgrids,
		ymin=1, ymax=20, ymajorgrids,
		zmin=0, zmax=1, zmajorgrids,
		axis lines=left,
		grid=major
		]
		%\addplot3[surf, z buffer=sort, colormap/jet, shader=flat] file {consistency_n_20.dat};
		%\addplot3[patch,patch refines=2,shader=faceted interp,patch type=biquadratic] file {consistency_n_10.dat};
		%\addplot3[patch,patch type=biquadratic, shader=faceted interp,patch refines=3] file {consistency_n_10.dat};
		\addplot3+[only marks,scatter] file {consistency_n_20.dat}; 
		%\addplot3+[mesh,scatter,samples=10,domain=1:20] file {consistency_n_20.dat}; 
	\end{axis}
\end{tikzpicture}
\caption{Consistency Level $\Phi$ at $N=20$.}
\label{fig:Phi}
\end{figure}
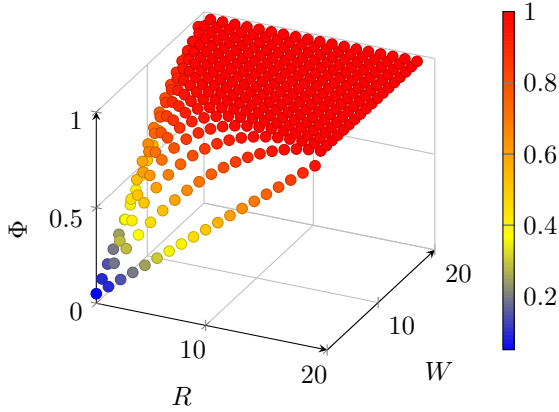

\noindent \textbf{Measuring the Consistency Level.}
We chose the probability that a read returns the most recent version as the consistency level indicator ($\Phi$) (shown in (\ref{eqn:Phi})). In case of strong consistency ($R + W \geq N$), $\Phi = 1$ otherwise $\Phi = 1 - p_{s}$ where $p_{s}$ (shown in (\ref{eqn:ps})) is the probability that the read quorum does not include the last up-to-date version \cite{bailis2012probabilistically}. Figure \ref{fig:Phi}, shows $\Phi$ versus $R$ and $W$ in case of $N = 20$. $R$, $W$ and $N$ are positive integer values ($\in Z^{+}$), hence $\Phi(R, W, R)$ is a discrete function.

\begin{align}
	p_{s} = \frac{\dbinom{N - W}{R}}{\dbinom{N}{R}} && \cite{bailis2012probabilistically}
	\label{eqn:ps}
\end{align}

\begin{equation}
	\Phi (R, W, N) =
		\begin{cases}
			1 - p_{s} & R + W \le N\\
			1         & R + W > N
		\end{cases}
		\label{eqn:Phi}
\end{equation}

\noindent \textbf{Controlling the Consistency Level.}
Once the adaptation module chooses a certain value ($\phi$) for ($\Phi$) that supposedly satisfies the application-specific performance indicator ($\chi$), the tunable consistency module needs to find values for $R$, $W$ and $N$ that gives ($\phi^{'} = \Phi(R, W, N)$), where ($\phi^{'}$) is as close as possible to the given $\phi$ (recall that \{$R$, $W$, $N$\} $\in Z^{+}$, and $\Phi(R, W, N)$ is a discrete function). We assume $N$ is constant per-application and is set as a system-wide policy by the network administrator. In (\ref{eqn:prop-proof}), we prove that swapping the values of $R$ and $W$ yields the same value for $\Phi$. This property helps in reducing the search space.

\begin{align*}
	\Phi(R, W, N) &= 1 - \frac{\frac{(N - W)!}{(N - W - R)! \times R!}}{\frac{N!}{(N - R)! \times R!}} \nonumber && {\scriptstyle(Case: R + W \le N)} \\
	            &= 1 -\frac{(N - W)! \times (N - R)!}{(N - W - R)! \times N!} \nonumber
\end{align*}

\begin{align}
	\Phi(W, R, N) &= 1 - \frac{(N - R)! \times (N - W)!}{(N - R - W)! \times N!} \nonumber \\
	                &= \Phi(R, W, N) && \blacksquare
	\label{eqn:prop-proof}
\end{align}

The values for $R$ and $W$ that gives the nearest value to a certain value $\phi$ for $\Phi(R, W, N)$ could be found using (\ref{eqn:phi2rw}). A simple algorithm is shown in Algorithm \ref{algo:phi2rw} for finding the values of $R$ and $W$.
 
\begin{equation}
	<R, W> = \argmin_{i,j} \Vert \Phi(i, j, N) - \phi \Vert
	\label{eqn:phi2rw}
\end{equation}

\begin{algorithm}[h]
	\small
	\DontPrintSemicolon
	\KwData{$N$, number of replicas}
	\Begin{
		min $\leftarrow$ $\infty$\;
		\For{$i \in [1, N)$}{
			\For{$j \in [i, N - i]$}{
				dist $\leftarrow$ $\Vert\Phi(i, j, N) - \phi\Vert$\;
				\If{dist < min}{
					min $\leftarrow$ dist\;
					R $\leftarrow$ i\;
					W $\leftarrow$ j\;
				}
			}
		}
	}
	\caption{Given a certain value ($\phi$) for ($\Phi$) find appropriate values for $R$ and $W$ in $\mathcal{O}(2)$.}
	\label{algo:phi2rw}
\end{algorithm}

\subsection{Adaptation Module}\label{sec:proposed:am}
The adaptation module is responsible for selecting an appropriate configuration (\emph{i.e.,} consistency level ($\Phi$)) for the tunable consistency module given a certain performance level ($\chi$) which is calculated with the help of the stored procedure compiler module. In this section, we show how clustering can be used by the adaptation module in order to map a certain performance level ($\chi$) into a corresponding consistency level ($\Phi$).

\noindent \textbf{Monitoring.}
In order for the adaptation module to function properly, it needs to continuously collect sample data about the application performance and configuration of the tunable consistency module. In particular, it collects different values for the consistency level indicator ($\Phi$) and notes the corresponding performance level ($\chi$), then uses these values to update the clustering technique.

\noindent \textbf{Clustering.}
We use clustering in order to find a mapping between the application performance indicator ($\chi$) and the consistency level ($\Phi$).
First, the collected data is clustered by the application performance indicator ($\chi$) and each center will be a consistency level ($\Phi$).
Next, when a specific level of application performance is needed, the nearest cluster to the required performance level is located and the value of the associated consistency level ($\Phi$) will be used to select appropriate values for $R$ and $W$. We opt for online incremental clustering techniques \cite{beringer2006online, haveliwala2000scalable}. Although such techniques can yield less accurate results compared to those offline techniques but they scale better in terms of storage and they do not require re-clustering with every new measurement. We tested two online techniques: (1) Sequential K-means, and (2) Incremental K-means.

\noindent \textbf{Re-Clustering.}
In oftentimes the adaptation module may need to recalculate the cluster heads.
This is needed when there is a change in the network that affects the accuracy of the adaptation module in finding the closest configuration for a given performance level.

\noindent \textbf{Sequential K-means Clustering.}
The first technique that we tested was the sequential K-means clustering (shown in Algorithm \ref{algo:kmeans-seq}). Algorithm \ref{algo:kmeans-seq} is our adoptation of the ``sequential K-means'' algorithm presented in \cite{ackerman2014incremental}. This technique requires the number of clusters to be initially specified. The first n-measurement will be assigned to the n-clusters, and then every new measurement will be assigned to the nearest cluster, and finally the cluster's mean will be updated.

\begin{algorithm}[h]
	\small
	\DontPrintSemicolon
	\SetKw{KwGoTo}{goto}
	\KwData{$\chi_{k}$, $k^{th}$ application's specific performance indicator}
	\KwData{$\Phi_{k}$, $k^{th}$ consistency level indicator}
	\KwData{$N_{c}$, number of clusters}
	\KwData{$N_{p}$, number of data points per cluster}
	\KwData{$N_{t}$, total number of data points}
	\SetKwFunction{nearest}{nearest}
	\Begin{
		$N_{t}$ $\leftarrow$ $\emptyset$\;
		\If {$N_{t}$ $<$ $N_{c}$} {
			$C_{k}.\chi$ $\leftarrow$ $\chi_{k}$\;
			$C_{k}.\Phi$ $\leftarrow$ $\Phi_{k}$\;
			$C_{k}.N_{p}$ $\leftarrow$ $C_{k}.N_{p}$ + $1$\;
		}
		\Else {
				$i_{c}$ = \nearest($\chi_{k}$, $C$)\;
				$C_{i_{c}}.\chi$ $\leftarrow$ $(C_{i_{c}}.\chi$ * $C_{i_{c}}.N_{p}) + \chi_{k}$\;
				$C_{i_{c}}.\Phi$ $\leftarrow$ $(C_{i_{c}}.\Phi$ * $C_{i_{c}}.N_{p}) + \Phi_{k}$\;
				
				$C_{i_{c}}.N_{p}$ $\leftarrow$ $C_{i_{c}}.N_{p}$ + 1\;
				
				$C_{i_{c}}.\chi$ $\leftarrow$ $C_{i_{c}}.\chi$ / $C_{i_{c}}.N_{p}$\;
				$C_{i_{c}}.\Phi$ $\leftarrow$ $C_{i_{c}}.\Phi$ / $C_{i_{c}}.N_{p}$\;
		}
		$N_{t}$ $\leftarrow$ $N_{t} + 1$\;
		\SetKwProg{nearest}{Function}{}{}
		\nearest{nearest}{
				\KwData{$d$, datapoint}
				\KwData{C$_{N}$, set of $N$ clusters}
				\Begin{
					idx $\leftarrow$ $\emptyset$; min $\leftarrow$ $\infty$\;
					\For{$i \in N$}{
						dist $\leftarrow$ $\Vert d.\chi - C_{i}.\chi \Vert$\;
						\If{dist $<$ min}{
							min $\leftarrow$ dist\;
							idx $\leftarrow$ i\;
						}
					}
					\KwRet idx\;
				}
		}
	}
	\caption{Using Sequential K-means Clustering at the Adaptation Module.}
	\label{algo:kmeans-seq}
\end{algorithm}

\noindent \textbf{Incremental K-means Clustering.}
The second technique that we tested was the incremental K-means clustering (shown in Algorithm \ref{algo:kmeans-incr}). We adopt the ``incremental clustering'' algorithm presented in \cite{rokach2005clustering} as a base for Algorithm \ref{algo:kmeans-incr}. This technique does not require the number of clusters to be specified as it uses a dynamic number of clusters. Every new measurement will be assigned to the nearest cluster if it is close enough (based on a threshold). If none was found, then a new cluster will be added that includes this measurement. The threshold depends on the performance indicator $\chi$, thus we use the relative error as the distance measure to allow the use of a single threshold value for different performance indicators.

\begin{algorithm}[h]
	\small
	\DontPrintSemicolon
	\SetKw{KwGoTo}{goto}
	\KwData{$\chi_{k}$, $k^{th}$ application's specific performance indicator}
	\KwData{$\Phi_{k}$, $k^{th}$ consistency level indicator}
	\KwData{$N_{c}$, number of clusters}
	\KwData{$N_{p}$, number of data points per cluster}
	%\KwData{$N_{t}$, total number of data points}
	\KwData{$\tau$, threshold}
	\SetKwFunction{nearest}{nearest}
	\Begin{
		%$N_{t}$ $\leftarrow$ $\emptyset$\;
		\If {$N_{c}$ $>$ $0$} {
			$i_{c}$ = \nearest($\chi_{k}$, $C$) %\tcp*{same as Algo. \ref{algo:kmeans-seq}}
			\If {$\Vert$ $C_{i_{c}}.\chi$ - $\chi_{k}$ $\Vert$ / $C_{i_{c}.\chi} < \tau$} {
				$C_{i_{c}}.\chi$ $\leftarrow$ $(C_{i_{c}}.\chi$ * $C_{i_{c}}.N_{p}) + \chi_{k}$\;
				$C_{i_{c}}.\Phi$ $\leftarrow$ $(C_{i_{c}}.\Phi$ * $C_{i_{c}}.N_{p}) + \Phi_{k}$\;
				
				$C_{i_{c}}.N_{p}$ $\leftarrow$ $C_{i_{c}}.N_{p}$ + 1\;
				
				$C_{i_{c}}.\chi$ $\leftarrow$ $C_{i_{c}}.\chi$ / $C_{i_{c}}.N_{p}$\;
				$C_{i_{c}}.\Phi$ $\leftarrow$ $C_{i_{c}}.\Phi$ / $C_{i_{c}}.N_{p}$\;
			}
			\Else {
				$C$.create\_new\_cluster($\chi_{k}$, $\Phi_{k}$)\;
			}
		}
		\Else {
			$C$.create\_new\_cluster($\chi_{k}$, $\Phi_{k}$)\;
		}
		\SetKwProg{nearest}{Function}{}{}
		\nearest{nearest}{
				\KwData{$d$, datapoint}
				\KwData{C$_{N}$, set of $N$ clusters}
				\Begin{
					idx $\leftarrow$ $\emptyset$; min $\leftarrow$ $\infty$\;
					\For{$i \in N$}{
						dist $\leftarrow$ $\Vert d.\chi - C_{i}.\chi \Vert$\;
						\If{dist < min}{
							min $\leftarrow$ dist\;
							idx $\leftarrow$ i\;
						}
					}
					\KwRet idx\;
				}
		}
	}
	\caption{Using Incremental K-means Clustering at the Adaptation Module.}
	\label{algo:kmeans-incr}
\end{algorithm}

\noindent \textbf{Latency.}
In some cases, a given performance level can be satisfied by a set of different $R$ and $W$ pairs.
Even though selecting any of them has no impact on the performance, it can have an impact on the latency.
The tunable consistency module will monitor the frequency of reads and writes for each application and given the property ($\Phi(W, R, N) = \Phi(R, W, N)$) proved in (\ref{eqn:prop-proof}).
If the application tends to do more reads then the tunable consistency module will set $R$ to be the smallest value in order to reduce the read latency, while if the application tends to do more writes then the tunable consistency module will set $W$ to be the smallest value in order to reduce the write latency.

\subsection{Application-Controller Interaction}

\begin{figure}[h]
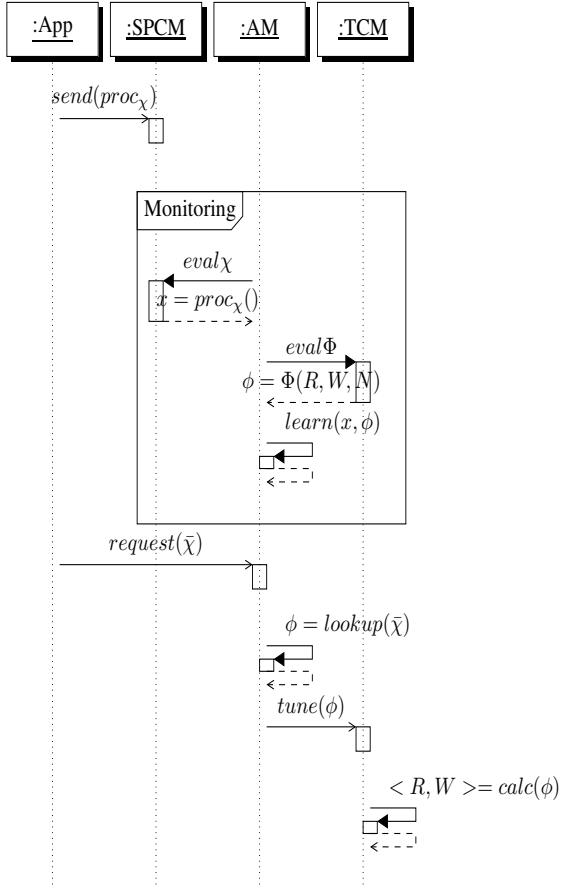

	\centering
	\resizebox{0.45\textwidth}{0.5\textheight}{
	\begin{sequencediagram}
		\newinst{app}{:App}
		\newinst{spcm}{:SPCM}
		\newinst{am}{:AM}
		\newinst{tcm}{:TCM}
		
		% init
		\begin{messcall}{app}{$send(proc_{\chi})$}{spcm}{}
		\end{messcall}
		
		% bootstrap
		\begin{sdblock}{Monitoring}{}
			\begin{call}{am}{$eval \chi$}{spcm}{$x = proc_{\chi}()$}
			\end{call}

			\begin{call}{am}{$eval \Phi$}{tcm}{$\phi = \Phi(R, W, N)$}
			\end{call}

			\begin{callself}{am}{$learn(x, \phi)$}{}
			\end{callself}
		\end{sdblock}

		% main		
		\begin{messcall}{app}{$request(\bar{\chi})$}{am}{}
		\end{messcall}

		\begin{callself}{am}{$\phi = lookup(\bar{\chi})$}{}
		\end{callself}

		\begin{messcall}{am}{$tune(\phi)$}{tcm}{}
		\end{messcall}

		\begin{callself}{tcm}{$<R, W> = calc(\phi)$}{}
		\end{callself}
	\end{sequencediagram}
	}
	\caption{The Sequence Diagram}
	\label{fig:seq-diag}
\end{figure}

Figure \ref{fig:seq-diag} shows a sequence diagram for the proposed adaptation strategy. It shows the interaction between the application (App) and the various controller modules: stored-procedure compiler module (SPCM), adaptation module (AM), and the tunable consistency module (TCM).
Initially, the application creates a stored procedure ($proc_{\chi}$) that is responsible for calculating the application-specific performance indicator ($\chi$), and then sends that procedure to the controller where it gets executed by the stored-procedure compiler module when needed.
Next, the controller monitors and gathers samples ($x$ and $\phi$) for the application-specific performance indicator ($\chi$) and the corresponding consistency level ($\Phi$), respectively. Then for each sample, the adaptation module invokes the clustering algorithm ($learn(x, \phi)$).
Finally, when the application notifies ($request(\bar{\chi})$) the controller with a desired value ($\bar{\chi}$) for the performance indicator ($\chi$), the adaptation module uses the clustering algorithm to find an estimate ($lookup(\bar{\chi})$) for a corresponding value ($\phi$) for the consistency level indicator ($\Phi$). Then, the adaptation module notifies ($tune(\phi)$) the tunable consistency module with this value, which in-turn calculates ($calc(\phi)$) the module internal parameters ($R$ and $W$) and applies such configuration.

\section{Evaluation}\label{sec:results}
In order to evaluate the validity of the proposed adaptation strategy, we evaluated the effectiveness of the clustering techniques (sequential and incremental) in mapping performance indicators ($\chi$) to consistency levels ($\Phi$). In our evaluation, we assumed that the relationship between the application-specific performance indicator ($\chi$) and the consistency level indicator ($\Phi$) is one the following relations: (1) linear ($\chi = A\Phi + C$), (2) quadratic ($\chi = A\Phi^2 + B\Phi + C$), (3) cubic ($\chi = A\Phi^3 + B\Phi^2 + C\Phi + D$), or (4) logarithmic ($\chi = A.log_{10}(\Phi) + C$). $A$, $B$, $C$, and $D$ are constants. We used a sample of 1000 uniform random numbers to bootstrap the algorithms. Then, we chose 100 arbitrary uniform random test values for $\chi$ and let the adaptation module figure out appropriate values for $\Phi$ that satisfies the given values for $\chi$, and calculate the RMSE between the given $\chi$ values and the ones calculated using values of $\Phi$ returned by the adaptation strategy.

Figure \ref{fig:seq-rmse-vs-clusters} shows the RMSE of the sequential K-means technique (Algorithm \ref{algo:kmeans-seq}) versus the number of clusters. The results show, in the cases we tested, that with a reasonable number of clusters ($\geq$ 50) a plausible mapping (low RMSE) could be estimated between the application performance indicators ($\chi$) which we tested and the consistency level indicator ($\Phi$).

\begin{figure}[h!]
\centering
%\resizebox{!}{0.20\textheight}{
\resizebox{!}{0.15\textheight}{
	\subfloat[RMSE vs Threshold]{
		\begin{tikzpicture}
		\begin{axis}[
			xlabel=Threshold,
			ylabel=RMSE,
			cycle list name=exotic,
			legend style={at={(0.02,0.98)},anchor=north west}
			]
			\legend{Linear, Quadratic, Cubic, Logarithmic}
			\addplot+[mark=star] table[x index=0, y index=2] {incr_linear.dat};
			\addplot+[mark=o] table[x index=0, y index=2] {incr_quadratic.dat};
			\addplot+[mark=square] table[x index=0, y index=2] {incr_cubic.dat};
			\addplot+[mark=diamond] table[x index=0, y index=2] {incr_log.dat};
		\end{axis}
		\end{tikzpicture}
	}
	\subfloat[Number of Cluster Heads vs Threshold]{
		\begin{tikzpicture}
		\begin{axis}[
			xlabel=Threshold,
			ylabel=Number of Cluster Heads,
			cycle list name=exotic
			]
			\legend{Linear, Quadratic, Cubic, Logarithmic}
			\addplot+[mark=star] table[x index=0, y index=1] {incr_linear.dat};
			\addplot+[mark=o] table[x index=0, y index=1] {incr_quadratic.dat};
			\addplot+[mark=square] table[x index=0, y index=1] {incr_cubic.dat};
			\addplot+[mark=diamond] table[x index=0, y index=1] {incr_log.dat};
		\end{axis}
		\end{tikzpicture}
	}

	%\begin{groupplot}[group style={group name=my plots,group size= 2 by 4,horizontal sep=40pt},width=4.5cm]

	%\nextgroupplot[title={Linear}, ylabel={RMSE}, ylabel absolute, ylabel style={yshift=-3ex}]
	%	\addplot[color=red, mark=x] table[x index=0, y index=2] {incr_linear.dat};
	%\nextgroupplot[title={Linear}, ylabel={Clusters}, ylabel absolute, ylabel style={yshift=-4ex}]
	%	\addplot[color=blue, mark=x] table[x index=0, y index=1] {incr_linear.dat};
		
	%\nextgroupplot[title={Quadratic}, ylabel={RMSE}, ylabel absolute, ylabel style={yshift=-3ex}]
	%	\addplot[color=red, mark=x] table[x index=0, y index=2] {incr_quadratic.dat};
	%\nextgroupplot[title={Quadratic}, ylabel={Clusters}, ylabel absolute, ylabel style={yshift=-4ex}]
	%	\addplot[color=blue, mark=x] table[x index=0, y index=1] {incr_quadratic.dat};
		
	%\nextgroupplot[title={Cubic}, ylabel={RMSE}, ylabel absolute, ylabel style={yshift=-3ex}]
	%	\addplot[color=red, mark=x] table[x index=0, y index=2] {incr_cubic.dat};
	%\nextgroupplot[title={Cubic}, ylabel={Clusters}, ylabel absolute, ylabel style={yshift=-4ex}]
	%	\addplot[color=blue, mark=x] table[x index=0, y index=1] {incr_cubic.dat};

	%\nextgroupplot[title={Logarithmic}, ylabel={RMSE}, ylabel absolute, ylabel style={yshift=-3ex}]
	%	\addplot[color=red, mark=x] table[x index=0, y index=2] {incr_log.dat};
	%\nextgroupplot[title={Logarithmic}, ylabel={Clusters}, ylabel absolute, ylabel style={yshift=-4ex}]
	%	\addplot[color=blue, mark=x] table[x index=0, y index=1] {incr_log.dat};
		
	%\end{groupplot}
%\end{tikzpicture}
}
\caption{Root Mean Square Error (RMSE) vs Threshold.}
\label{fig:incr-rmse-vs-threshold}
\end{figure}
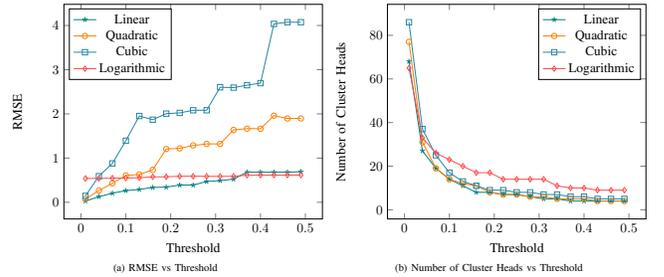

Figure \ref{fig:incr-rmse-vs-threshold} shows the RMSE of the incremental K-means technique (Algorithm \ref{algo:kmeans-incr}) versus the threshold. The results show, in the cases we tested, that a plausible mapping (low RMSE) could be estimated using a reasonable number of clusters ($\geq$ 50) by using a relatively small threshold ($\simeq$ 0.01). 

The results also indicate that even though online clustering techniques can yield less accurate results compared to offline techniques however in the cases we tested the online clustering techniques were sufficient with a reasonable number of clusters being used.

\begin{figure}[h!]
\centering
\resizebox{!}{0.24\textheight}{
%\resizebox{!}{0.15\textheight}{
\begin{tikzpicture}
	\begin{axis}[
			xlabel=Number of Clusters Heads,
			ylabel=RMSE,
			cycle list name=exotic
		]
		\legend{Linear, Quadratic, Cubic, Logarithmic}
		\addplot+[mark=star] file {linear.dat};
		\addplot+[mark=o] file {quadratic.dat};
		\addplot+[mark=square] file {cubic.dat};
		\addplot+[mark=diamond] file {log.dat};
	\end{axis}
	%\begin{groupplot}[group style={group name=my plots,group size= 2 by 2},width=4.5cm]

	%\nextgroupplot[title={Linear}, ylabel={RMSE}]
		%\addplot[color=red,mark=x] file {linear.dat};
	
	%\nextgroupplot[title={Quadratic}]
		%\addplot[color=red,mark=x] file {quadratic.dat};
		
	%\nextgroupplot[title={Cubic}, ylabel={RMSE}]
		%\addplot[color=red,mark=x] file {cubic.dat};

	%\nextgroupplot[title={Logarithmic}]
		%\addplot[color=red,mark=x] file {log.dat};
		
	%\end{groupplot}
\end{tikzpicture}
}
\caption{Root Mean Square Error (RMSE) vs Number of Cluster Heads.}
\label{fig:seq-rmse-vs-clusters}
\end{figure}
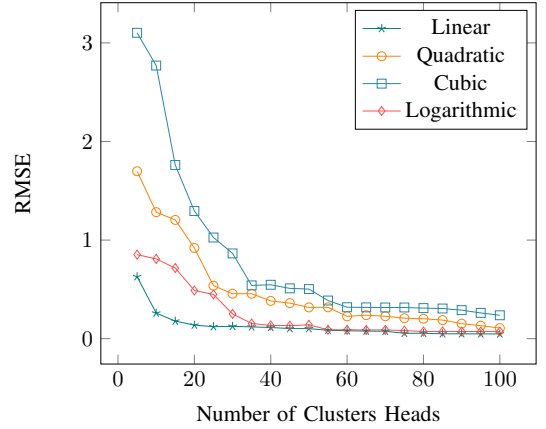

\section{Conclusion and Future Work}\label{sec:conclusion}

In this paper, we examined the feasibility of using adaptive controllers that are built on-top of tunable consistency models similar to that of Apache Cassandra.
We presented an adaptation strategy that selects feasible values for the consistency level indicator ($\Phi$) that satisfies a given application performance indicator ($\chi$).
We employed two online clustering techniques (sequential and incremental K-means) in order to find suitable mapping between $\chi$ and $\Phi$.
In the cases that we tested, our results showed that in the case of sequential K-means, with a reasonable number of clusters ($\geq$ 50), a plausible mapping (low RMSE) could be estimated between the application performance indicators ($\chi$) and the consistency level indicator ($\Phi$). In the case of incremental K-means, the results also showed that a plausible mapping (low RMSE) could be estimated using a similar number of clusters ($\geq$ 50) by using a small threshold ($\simeq$ 0.01).

In the future, we plan to evaluate the validity and effectiveness of the proposed consistency adaptation strategy using an implementation of an SDN application running on top of a cluster of a distributed controllers.

\section{Acknowledgments}
The second author acknowledge support from the Natural Sciences and Engineering Research Council of Canada (NSERC) through the NSERC Discovery Grant program.

\bibliographystyle{IEEEtran}
\bibliography{references}

\begin{thebibliography}{10}
\providecommand{\url}[1]{#1}
\csname url@samestyle\endcsname
\providecommand{\newblock}{\relax}
\providecommand{\bibinfo}[2]{#2}
\providecommand{\BIBentrySTDinterwordspacing}{\spaceskip=0pt\relax}
\providecommand{\BIBentryALTinterwordstretchfactor}{4}
\providecommand{\BIBentryALTinterwordspacing}{\spaceskip=\fontdimen2\font plus
\BIBentryALTinterwordstretchfactor\fontdimen3\font minus
  \fontdimen4\font\relax}
\providecommand{\BIBforeignlanguage}[2]{{%
\expandafter\ifx\csname l@#1\endcsname\relax
\typeout{** WARNING: IEEEtran.bst: No hyphenation pattern has been}%
\typeout{** loaded for the language `#1'. Using the pattern for}%
\typeout{** the default language instead.}%
\else
\language=\csname l@#1\endcsname
\fi
#2}}
\providecommand{\BIBdecl}{\relax}
\BIBdecl

\bibitem{aslan2016adaptive}
M.~Aslan and A.~Matrawy, ``{Adaptive Consistency for Distributed SDN
  Controllers},'' in \emph{Proceedings of the 17th International Network
  Strategy and Planning Symposium (Networks 2016)}, 2016,
  \url{http://www.sce.carleton.ca/~maslan/files/sdn-adaptive.pdf}.

\bibitem{dixit2013towards}
A.~Dixit, F.~Hao, S.~Mukherjee, T.~Lakshman, and R.~Kompella, ``Towards an
  elastic distributed sdn controller,'' in \emph{Proceedings of the second ACM
  SIGCOMM workshop on Hot topics in software defined networking}.\hskip 1em
  plus 0.5em minus 0.4em\relax ACM, 2013, pp. 7--12.

\bibitem{tootoonchian2010hyperflow}
A.~Tootoonchian and Y.~Ganjali, ``Hyperflow: A distributed control plane for
  openflow,'' in \emph{Proceedings of the 2010 internet network management
  conference on Research on enterprise networking}.\hskip 1em plus 0.5em minus
  0.4em\relax USENIX Association, 2010, pp. 3--3.

\bibitem{levin2012logically}
D.~Levin, A.~Wundsam, B.~Heller, N.~Handigol, and A.~Feldmann, ``Logically
  centralized?: state distribution trade-offs in software defined networks,''
  in \emph{Proc. of the first workshop on Hot topics in software defined
  networks}.\hskip 1em plus 0.5em minus 0.4em\relax ACM, 2012, pp. 1--6.

\bibitem{koponen2010onix}
T.~Koponen, M.~Casado, N.~Gude, J.~Stribling, L.~Poutievski, M.~Zhu,
  R.~Ramanathan, Y.~Iwata, H.~Inoue, T.~Hama \emph{et~al.}, ``Onix: A
  distributed control platform for large-scale production networks.'' in
  \emph{OSDI}, vol.~10, 2010, pp. 1--6.

\bibitem{berde2014onos}
P.~Berde, M.~Gerola, J.~Hart, Y.~Higuchi, M.~Kobayashi, T.~Koide, B.~Lantz,
  B.~O'Connor, P.~Radoslavov, W.~Snow \emph{et~al.}, ``Onos: towards an open,
  distributed sdn os,'' in \emph{Proceedings of the third workshop on Hot
  topics in software defined networking}.\hskip 1em plus 0.5em minus
  0.4em\relax ACM, 2014, pp. 1--6.

\bibitem{panda2013cap}
A.~Panda, C.~Scott, A.~Ghodsi, T.~Koponen, and S.~Shenker, ``Cap for
  networks,'' in \emph{Proc. of the second ACM SIGCOMM workshop on Hot topics
  in software defined networking}.\hskip 1em plus 0.5em minus 0.4em\relax ACM,
  2013, pp. 91--96.

\bibitem{brewer2000towards}
E.~Brewer, ``Towards robust distributed systems,'' in \emph{PODC}, 2000, p.~7.

\bibitem{brewer2012cap}
------, ``Cap twelve years later: How the ``rules'' have changed,''
  \emph{Computer}, vol.~45, no.~2, pp. 23--29, 2012.

\bibitem{cassandraconfig}
(2016) {Apache Cassandra: configuring data consistency}.
  \url{http://docs.datastax.com/en/cassandra/2.0/cassandra/dml/dml_config_consistency_c.html}.

\bibitem{yu2000building}
H.~Yu and A.~Vahdat, ``Building replicated internet services using tact: A
  toolkit for tunable availability and consistency tradeoffs,'' in
  \emph{Advanced Issues of E-Commerce and Web-Based Information Systems, 2000.
  WECWIS 2000. Second International Workshop on}.\hskip 1em plus 0.5em minus
  0.4em\relax IEEE, 2000, pp. 75--84.

\bibitem{Guo201495}
Z.~Guo, M.~Su, Y.~Xu, Z.~Duan, L.~Wang, S.~Hui, and H.~J. Chao, ``Improving the
  performance of load balancing in software-defined networks through load
  variance-based synchronization,'' \emph{Computer Networks}, vol.~68, no.~0,
  pp. 95 -- 109, 2014, communications and Networking in the Cloud.

\bibitem{aslan2016impact}
M.~Aslan and A.~Matrawy, ``On the impact of network state collection on the
  performance of sdn applications,'' \emph{IEEE Communications Letters},
  vol.~20, no.~1, pp. 5--8, 2016.

\bibitem{lakshman2009cassandra}
A.~Lakshman and P.~Malik, ``Cassandra: structured storage system on a p2p
  network,'' in \emph{Proceedings of the 28th ACM symposium on Principles of
  distributed computing}.\hskip 1em plus 0.5em minus 0.4em\relax ACM, 2009, pp.
  5--5.

\bibitem{lakshman2010cassandra}
------, ``Cassandra: a decentralized structured storage system,'' \emph{ACM
  SIGOPS Operating Systems Review}, vol.~44, no.~2, pp. 35--40, 2010.

\bibitem{sivasubramanian2012amazon}
S.~Sivasubramanian, ``Amazon dynamodb: a seamlessly scalable non-relational
  database service,'' in \emph{Proceedings of the 2012 ACM SIGMOD International
  Conference on Management of Data}.\hskip 1em plus 0.5em minus 0.4em\relax
  ACM, 2012, pp. 729--730.

\bibitem{bailis2012probabilistically}
P.~Bailis, S.~Venkataraman, M.~J. Franklin, J.~M. Hellerstein, and I.~Stoica,
  ``Probabilistically bounded staleness for practical partial quorums,''
  \emph{Proceedings of the VLDB Endowment}, vol.~5, no.~8, pp. 776--787, 2012.

\bibitem{iso:sql}
``{Information technology -- Database languages -- SQL -- Part 4: Persistent
  Stored Modules (SQL/PSM)},'' International Organization for Standardization,
  Geneva, CH, Standard, 2011.

\bibitem{beringer2006online}
J.~Beringer and E.~H{\"u}llermeier, ``Online clustering of parallel data
  streams,'' \emph{Data \& Knowledge Engineering}, vol.~58, no.~2, pp.
  180--204, 2006.

\bibitem{haveliwala2000scalable}
T.~Haveliwala, A.~Gionis, and P.~Indyk, ``Scalable techniques for clustering
  the web,'' 2000.

\bibitem{ackerman2014incremental}
M.~Ackerman and S.~Dasgupta, ``Incremental clustering: The case for extra
  clusters,'' in \emph{Advances in Neural Information Processing Systems},
  2014, pp. 307--315.

\bibitem{rokach2005clustering}
L.~Rokach and O.~Maimon, ``Clustering methods,'' in \emph{Data mining and
  knowledge discovery handbook}.\hskip 1em plus 0.5em minus 0.4em\relax
  Springer, 2005, pp. 321--352.

\end{thebibliography}
\end{document}